\documentstyle[prd,tighten,aps]{revtex}
\begin{document} 
\draft 
\twocolumn[\hsize\textwidth\columnwidth\hsize\csname
@twocolumnfalse\endcsname
\begin{flushright}
Preprint DFPD 96/TH/58\\ 
hep-th/9611100 \\
November, 1996 
\end{flushright}
\title{On Lorentz Invariant Actions for Chiral
P--Forms}
\author{
Paolo Pasti$^1$, 
Dmitri Sorokin$^{1,2}$ and 
Mario Tonin$^1$
}
\address{${}^1$ Universit\`a Degli Studi Di Padova,
Dipartimento Di Fisica ``Galileo Galilei''\\
ed INFN, Sezione Di Padova,
Via F. Marzolo, 8, 35131 Padova, Italia\\
${}^2$ on leave from Kharkov Institute of Physics and Technology, 
Kharkov 310108, Ukraine} 
\maketitle 
\begin{abstract} 
We demonstrate how a Lorentz covariant formulation of the 
chiral p-form model in $D=2(p+1)$
containing infinitely many auxiliary fields is related to a Lorentz 
covariant formulation with only one auxiliary scalar field entering a 
chiral p--form action in a nonpolynomial way. The latter can be regarded 
as a consistent Lorentz--covariant truncation of the former.
We make the Hamiltonian analysis of the model based on 
the nonpolynomial action and show that the Dirac constraints have 
a simple form and are all of the first class. In contrast to the Siegel 
model the constraints are not the square of second--class constraints.
The canonical Hamiltonian is quadratic and determines energy of
a single chiral p--form. 
In the case of $d=2$ chiral scalars the
constraint can be improved by use of a `twisting' procedure 
(without the loss of the property to be of the first class) in such a 
way that the central charge of the quantum constraint algebra is zero.
This points to possible absence  
of anomaly in an appropriate quantum version of the model.
\end{abstract}
  
\pacs{11.15-q, 11.17+y}

\vskip2pc]
\section{Introduction}
Chiral p--forms, i.e. antisymmetric boson fields with self--dual 
(p+1)--form field strengths, form integral part and play an 
important role in many 
theoretical models such as $D=6$ and type IIB $D=10$ supergravity, 
heterotic strings \cite{gsw} and M--theory five--branes (\cite{m,w,hs,js}
and references there in). A 
particular feature of these fields is that, since the self--duality 
condition implies the fulfillment of first--order equations of motion, 
which puts the theory on the mass shell, there is a problem of 
construction manifestly Lorentz invariant actions for the chiral
p--forms \cite{ms} and, as a consequence, a problem of quantizing such 
fields. The analogous problems exist in manifestly electric--magnetic 
duality formulation of $D=4$ Maxwell theory \cite{bs}, where the Maxwell 
field can be considered as a complex chiral two--form.

Non manifestly covariant actions were proposed for $d=2$ chiral scalars in 
\cite{fj}, for $D=4$ duality symmetric Maxwell fields in \cite{z,d}, for 
$D=2(p+1)$ chiral p--forms in \cite{hen} and for duality symmetric 
fields in space--time of any dimension in \cite{ss}. All of these 
actions lead to second class constraints on the chiral boson phase
space, which complicates the quantization procedure.

In \cite{si} a 
$D=2(p+1)$ Lorentz invariant action for chiral p--forms was constructed 
by squaring the second--class constraints and introducing first--class 
constraints thus obtained into the action with Lagrange multipliers. 
However, though the Lagrange multipliers do not contribute to the 
equations of motion of this model, it is not clear whether in $D> 2$
($p> 1)$ there is enough local symmetry to completely gauge them 
away \cite{si}. At the same time even in $d=2$ the Siegel action 
for chiral scalars is not easy to quantize (in particular because of an
anomaly problem) and an extensive literature has been devoted to 
studying this point (see, for example, \cite{is}).

Another covariant (Hamiltonian) formulation was proposed 
for $d=2$ chiral scalars by McClain, Wu and Yu \cite{mwy} (see also 
\cite{wot}) and 
generalized to the case of higher order chiral p--forms in \cite{mr,dh}. 
The construction is based on a procedure of converting the second--class 
constraints into first--class ones by introducing auxiliary fields 
\cite{fs}. In the case at hand this required an infinite set of 
auxiliary $(p+1)$--forms. By use of a Legendre transformation it is 
possible to write down a manifestly Lorentz invariant form of the chiral 
boson actions \cite{bk}. The chiral scalar and free Maxwell theory were 
consistently quantized in such a formulation, respectively in \cite{mwy} 
and \cite{mr}.

It is of interest and somehow indicative that for a chiral 4-form in ten 
dimensions the Lorentz covariant formulation of \cite{mwy,wot,mr,dh,bk} was, 
actually independently, derived from type IIB closed superstring field 
theory in \cite{ber1,ber2}. 

The infinite set of auxiliary fields in the chiral boson models requires 
caution to deal with when one studies equations of motion, makes 
Hamiltonian analysis, imposes admissible gauge--fixing conditions 
and quantizes 
the models \cite{mwy}-\cite{ber2}, since, in particular, 
this infinite set corresponds to the infinite number of local 
symmetries and first--class constraints which cause problems with 
choosing the right regularization procedure. For instance in \cite{mwy} 
a strong group--theoretical argument based on the existence of a 
symmetry of the quantum theory was used to justify the regularization 
which leads to the correct partition function of the chiral scalar.

Note also that a direct cutting of the infinite series of fields at a
number of N results in an action which does not describe a single chiral 
p-form \cite{bk}.

An alternative Lorentz invariant action for chiral p--forms was proposed 
in \cite{pst1,pst2,pst3}. This formulation involves finite number 
of auxiliary 
fields and, as a consequence, a finite number of local symmetries being 
sufficient
to gauge these fields away. 
Upon an appropriate gauge fixing one gets non--manifestly covariant 
models of Refs.\cite{fj,hen,ss,js}. 
The advantage of the covariant approach is 
that one should not bother about proving Lorentz invariance which may be 
rather cumbersome \cite{hen,ss,js}.
 
A minimal version of this covariant formulation contains
(in space--time of any even dimension \footnote{Remember that if $p$ is odd 
the chiral 
form is complex in $D=2(p+1)$}) only one scalar auxiliary field 
entering the action in a nonpolynomial way. In the case of $D=4$ Maxwell 
theory this scalar field was assumed to be of an axion nature 
\cite{pst2}.

The purpose of the present paper is, on the one hand, to show how 
the McClain--Wu--Yu approach and the approach of Refs. 
\cite{pst1,pst2,pst3} 
relate to each other, and, on the other hand, to make the Hamiltonian 
analysis of the nonpolynomial version and to demonstrate that in spite of 
the nonpolynomiality the structure of the constraints (which all belong 
to the first class) is rather simple, the canonical Hamiltonian is 
quadratic and describes the energy of a single chiral p--form boson.
In the case of a chiral scalar in $d=2$ the form of the first--class 
constraint allows improvement by ``twisting" its auxiliary field term, which 
at the quantum level allows the central charge of the constraint 
algebra to be zero. This points to possible absence of anomaly in a
quantum version of the model.

The paper is organized as follows. In Section 2 we review the Lorentz 
invariant chiral form actions of Refs. \cite{mwy}--\cite{ber2} and 
\cite{pst1,pst2,pst3} and demonstrate a relationship between them by 
either trying to get rid of the nonpolynomiality and eliminate the 
scalar auxiliary field at the expense of introducing auxiliary 
(p+1)--forms, or, vice versa, by a consistent truncation of the 
McClain--Wu--Yu infinite tail with putting on its end the auxiliary 
scalar field.
In Section 3 we analyse the classical Hamiltonian structure of the 
chiral form model with the single auxiliary scalar, and, in the $d=2$ 
case, discuss the problem of quantum anomaly of local symmetry 
of the model. 

In Conclusion open problems and prospectives are discussed.

To simplify notation and convention we consider $d=2$ chiral scalars 
and $D=6$ chiral 2--forms. However upon fitting numerical coefficients 
one can straightforwardly generalize all the expressions obtained to the 
generic case of chiral p--forms. We use almost positive signature of 
space--time, i.e. ($-,+,...,+)$. Latin letters stand for space--time 
indices $(l,m,n...=0,1,...,D-1)$ and Greek letters are spacial indices 
running from 1 to $D-1$.

\section{Relationship between Lorentz invariant chiral form actions}
\subsection{The infinite series action}
In a reduced form considered in \cite{wot,ber2}
(where part of an infinite number of auxiliary fields 
were eliminated by gauge fixing an infinite number of local symmetries)
the chiral boson 
action of \cite{mwy}--\cite{bk} in $D=6$ looks as follows:
$$
S=\int d^{6}x 
[-{1\over 6} F_{pqr}F^{pqr}+
\Lambda_{(1)}^{pqr}F_{pqr}
$$
\begin{equation}\label{inf}
-\sum_{n=0}^\infty (-1)^n
\Lambda_{(n+1)}^{pqr} \Lambda_{(n+2)pqr}].
\end{equation}
where $F_{pqr}=\partial_{[p}A_{qr]}$, and \\ 
$\Lambda_{(n+1)pqrs}$ =
${{(-1)^n}\over{3!}}\varepsilon_{pqrlmn}\Lambda^{lmn}_{(n+1)}$ 
form an infinite set
of (anti)--self--dual auxiliary three--form fields.
The action (\ref{inf}) describes a single physical chiral two--form $A_{mn}$ 
satisfying the selfduality condition:
\begin{equation}\label{sd}
{\cal F}^{lmn}\equiv
F_{lmn}-{1\over{3!}}\varepsilon_{lmnpqr}F^{pqr}=0.
\end{equation}
To arrive at the equation (\ref{sd}) one should make an assumption that
allowable are only those solutions to the equations of motion 
derived from (\ref{inf})   
which contain only a finite number of nonzero fields $\Lambda_{(n+1)}$.
This restriction, though it looks somewhat artificial, ensures the 
energy of the model to be well defined. 
Note that one cannot make such a truncation and eliminate all fields 
with $n$ greater than a given number $N$ directly in the action  
since this results in a model which does not describe a single chiral 
field, but an ordinary (chiral plus antichiral) antisymmetric gauge field, 
or a pair of chiral forms depending on the parity of $N$.
The reader may find a detailed
analysis of the model in \cite{dh,bk,ber2}.

\subsection{Chiral form action with a finite number of auxiliary fields}
The Lorentz invariant self--dual action of Refs. \cite{pst1,pst2,pst3} 
have the 
following form in $D=6$:
$$
S=\int d^6x [-{1\over 6}F_{lmn}F^{lmn}
+{1\over{2(u_qu^q)}}u^m{\cal F}_{mnl}{\cal F}^{nlr}u_r
$$
\begin{equation}\label{fine}
-\varepsilon^{mnpqrs}u_m\partial_n\Lambda_{pqrs}].
\end{equation}
Eq. (\ref{fine}) contains the anti--self--dual three--form ${\cal F}_{mnl}$  
defined in (\ref{sd}) (whose turning to zero on the mass shell results 
in the self--duality of $F_{lmn}$); an auxiliary vector field $u_m(x)$ and 
a four-form field $\Lambda_{lmnp}$.

 The action (\ref{fine}) is invariant under the following local  
transformations:
\begin{equation}\label{ordi}
\delta A_{mn}=\partial_{[m}\phi_{n]}(x),\qquad
\delta \Lambda_{lmnp}=\partial_{[l}\phi_{mnp]}(x)
\end{equation}
(which are the ordinary gauge symmetries of massless antisymmetric fields),
\begin{equation}\label{varm}
\delta A_{mn}={1\over 2}u_{[m}\varphi_{n]}(x), \qquad
\delta\Lambda_{lmnp}={1\over{u^2}}\varphi_{[l}u_m{\cal F}_{np]q}u^q,
\end{equation}
$$
\delta u_m=0
$$
(where $u^2=u_\alpha u_\alpha-u_0u_0$), and
\begin{equation}\label{var}
\delta u_m=\partial_m\varphi(x), \qquad
\delta A_{mn}=
{{\varphi(x)}\over{u^2}}{\cal F}_{mnp}u^p, 
\end{equation}
$$
\delta\Lambda_{mnpq}=-{{\varphi(x)}\over{(u^2)^2}}u^r{\cal F}_{r[mn}
{\cal F}_{pq]s}u^s.
$$

Note that the transformations (\ref{ordi}) and (\ref{varm}) are 
finite--step reducible, which is harmless. For instance, if 
in (\ref{varm}) $\varphi_m=\phi(x)u_m$ then the variations of $A_{mn}$ 
and $\Lambda_{mnpq}$ are zero.

The equation of motion of $\Lambda_{mnpq}$ reads
\begin{equation}\label{u}
{{\delta S}\over{\delta\Lambda}}~~~\Rightarrow~~~ \partial_{[m}u_{n]}=0.
\end{equation}
Its general solution is 
\begin{equation}\label{utoa}
u_m=\partial_ma(x)
\end{equation}
with $a(x)$ being a scalar field. Under (\ref{var}) $a(x)$ transforms as 
a Goldstone field ($\delta a(x)=\varphi(x)$) and can be completely gauge 
fixed. Thus $u_m$ is an auxiliary field. 
When one takes $u_m$ to be a unit time--like vector
(for instance $u_m=\delta^0_m$)
the model loses manifest Lorentz invariance and reduces 
to the noncovariant model of Refs. \cite{fj,hen,ss}.
If instead one chooses a space--like gauge
$u_m=\delta^5_m$ the action (\ref{fine}) 
reproduces the free chiral field formulation of Ref. \cite{js}.

We should note that because of 
the presence of the norm of $u_m$ in the denominator in 
the action (\ref{fine}) the gauge fixing condition $u_m=0$ 
$(a(x)=const)$ (or more generally $u_mu^m=0$) cannot be applied 
directly and in this sense is inadmissible. So in what follows we shall 
require  $u_mu^m\not = 0$. This situation is analogous 
to that in gravity, where one requires the existence of the inverse
space--time metric.
However, in principle, one can arrange a consistent limit of $u_m\to 0$
with an appropriate simultaneous limits of other fields in such a way 
that the physical contents of the model is the same as at other gauge 
points. 
\footnote{The problem of an admissible gauge choice also exists
for the infinitely--many--field actions \cite{ber2}. There it is caused by a 
requirement of convergency of infinite series. It might happen that
such ``critical'' gauge points in both approaches have a unique nature.}

The equation of motion of $A_{mn}$ is:
\begin{equation}\label{ea}
\epsilon^{lmnpqr}\partial_n({1\over{u^2}}u_p{\cal F}_{qrs}u^s)=0.
\end{equation}
Its general solution has the following form (when (\ref{u}) is taken 
into account):
\begin{equation}\label{gs}
{\cal 
F}_{lmn}u^n=u^2\partial_{[l}\Phi_{m]}+u^n(\partial_n\Phi_{[l})u_{m]}+
u_{[l}(\partial_{m]}\Phi_n)u^n,
\end{equation}
where $\Phi_m(x)$ is an arbitrary vector function.
One can check that the r.h.s. of (\ref{gs}) has the same form as the 
transformation of ${\cal F}_{qrs}u^s$ under (\ref{varm}), thus one can 
use this symmetry to gauge fix the r.h.s. of (\ref{gs}) to zero. As a 
result, because of the anti--self--duality, the whole ${\cal F}_{lmn}$ 
becomes equal to zero and we get the self--duality of ${F}_{lmn}$ 
(\ref{sd}) \cite{ss,pst1,pst2,pst3}. In this gauge the equation of motion of 
$u_m$ reduces to:
\begin{equation}\label{L}
{{\delta S}\over{\delta u}}~~~\Rightarrow~~~
\varepsilon^{mnpqrs}\partial_n\Lambda_{pqrs}=0,
\end{equation}
from which, in view of the local symmetry (\ref{ordi}), it follows that
$\Lambda_{mnpq}$ has only pure gauge degrees of freedom.

Thus the model based on action (\ref{fine}) indeed describes the 
classical dynamics of a single chiral two--form field $A_{mn}$.

We can simplify this action by substituting $u_m$ with its expression in 
terms of $a(x)$ (\ref{utoa}). Then (\ref{fine}) takes the form
which contains only one auxiliary scalar field $a(x)$
$$
S=\int d^6x [-{1\over 6}F_{lmn}F^{lmn}
$$
\begin{equation}\label{fina}
+{1\over{2(\partial_qa\partial^qa)}}\partial^ma(x){\cal F}_{mnl}
{\cal F}^{nlr}\partial_ra(x)].
\end{equation}
This action possesses the same symmetries as (\ref{fine}) 
with only difference that now $u_m=\partial_ma$, and $\Lambda_{lmnp}$
is absent from (\ref{ordi})-(\ref{var}). Notice that the variation  of 
the action (\ref{fina}) over $a(x)$ is identically zero on the solutions
(\ref{gs}) of Eq. (\ref{ea}). It is simply (\ref{ea}) multiplied by
${1\over{u^2}}{\cal F}_{lmf}u^f$ and, hence, does not produce new field 
equations. This reflects the presence of the local symmetry (\ref{var}). 

\subsection{Passing from one action to another}
Now let us try to relate action (\ref{fina}) to the action (\ref{inf})
containing infinite number of auxiliary fields. For this we should first get 
rid of the nonpolynomiality of (\ref{fina}) or (\ref{fine}) by 
introducing new auxiliary three--form fields in an appropriate way.
\footnote{Another way to eliminate nonpolynomiality 
is to consider $u_m$ to be a unit--norm harmonic--like variable, 
i.e. to impose the constraint $u^2=-1$. Such a version of the model 
was discussed in \cite{pst1,pst2}.}
We write:
$$
S=\int d^6x [-{1\over 6}F_{lmn}F^{lmn}
+{\hat\Lambda}_{(1)mnl}
{\cal F}^{mnl}
$$
\begin{equation}\label{final}
-{1\over 2}{\hat\Lambda}_{(1)lmn}{\hat\Lambda}_{(1)}^{lmn}+
{\hat\Lambda}^{lmn}_{(2)}({\hat\Lambda}_{(1)lmn}
-{\hat\Lambda}_{(0)lmp}\partial^pa\partial_na)].
\end{equation}
One can directly check that upon eliminating the auxiliary fields 
${\hat\Lambda}$ by solving their algebraic equations of motion one returns 
back to the action (\ref{fina}). 

We can make another step and replace
the term ${\hat\Lambda}_{(0)lmp}\partial^pa\partial_na$ in (\ref{final})
with an arbitrary three form $\Lambda_{(3)mnl}$ and, for not spoiling the 
model, add to the action one more term of the form
$$\int dx^6
{\hat\Lambda}^{lmn}_{(4)}({\hat\Lambda}_{(3)lmn}
-{\hat\Lambda}_{(0)lmp}\partial^pa\partial_na).
$$
Introducing more and more auxiliary three forms we can make any 
number N of steps of this kind and push the term containing $a(x)$ as far 
from the beginning of the series under construction as we like: 
$$
S=\int d^6x [-{1\over 6}F_{lmn}F^{lmn}
+{\hat\Lambda}_{(1)mnl}{\cal F}^{mnl}
$$
$$
-{1\over 2}{\hat\Lambda}_{(1)lmn}{\hat\Lambda}_{(1)}^{lmn}
+\sum_{n=0}^{2N-1} (-1)^n
{\hat\Lambda}_{(n+1)}^{pqr}{\hat\Lambda}_{(n+2)pqr}
$$
\begin{equation}\label{N}
+{\hat\Lambda}^{lmn}_{(2N+2)}({\hat\Lambda}_{(2N+1)lmn}
-{\hat\Lambda}_{(0)lmp}\partial^pa\partial_na)]
\end{equation}
At $N\to\infty$ 
we get exactly the action (\ref{inf}) upon splitting 
${\hat\Lambda}_{(n+1)pqr}$ $(n=0,...,2N+1)$ into self--dual and 
anti--self--dual parts and redefining them and their number 
in an appropriate way.

On the other hand if we start from the action (\ref{inf}) with the 
infinite number of fields, the procedure considered above prompts how one 
can consistently truncate the infinite series without spoiling the 
physical contents of the model at least at the classical level. The 
prescription is as follows: if in (\ref{inf}) one wants to put to zero 
all $\Lambda_{(n+1)}$ with $n>N'$ then one should replace the sum of 
the self--dual and anti--self--dual form 
$\Lambda_{(N')lmn}+\Lambda_{(N'+1)lmn}$ 
with a term of the form 
$\partial^pa{\hat\Lambda}_{(0)p[lm}\partial_{n]}a$ (where 
${\hat\Lambda_0}$ 
is an arbitrary three form).

Thus the chiral form action with infinite number of auxiliary fields is
related to the action (\ref{fina}) through the consistent truncation 
of the infinite tail of the former. The truncation leads to a
reconstruction of symmetries in the model which become
of the type written in Eqs. (\ref{ordi})--(\ref{var}).

\section{Hamiltonian analysis of the nonpolynomial action}
The Hamiltonian structure which follows from 
the chiral form action with infinite number of fields was discussed 
in detail in \cite{mwy,dh,bk} and we refer the reader to these papers.

Below we shall make the Hamiltonian analysis of models based on 
action (\ref{fina}).
As an instructive example we start with the action for a chiral boson 
in $d=2$ and compare its Hamiltonian structure with  
other versions \cite{si,mwy,wot} of the chiral boson model.

\subsection{$d=2$ chiral bosons}
Action (\ref{fina}) takes the following form \cite{pst3}:
\begin{equation}\label{d2}
S=\int d^2x{1\over 2}[\partial_{+}\phi\partial_{-}\phi-
{{\partial_{+}a}\over{\partial_{-}a}}
(\partial_{-}\phi)^2],
\end{equation}
where $\partial_\pm\equiv\partial_0\pm\partial_1$.

And the action--invariance transformations (\ref{ordi})--(\ref{var})  
reduce to
\begin{equation}\label{2}
\delta a=\varphi, \qquad \delta 
\phi={\varphi\over{\partial_-a}}\partial_-\phi.
\end{equation}

The essential difference of the action (\ref{d2}) from the Siegel model 
\cite{si} is that the second term in (\ref{d2}) contains derivatives of 
the scalar field $a(x)$ and not an arbitrary Lagrange multiplier 
$\lambda_{++}(x)$ as in the Siegel case.

The canonical momenta of the fields $\phi(x)$ and $a(x)$ are:
\begin{equation}\label{phi}
P_\phi={{\delta L}\over{\delta\dot\phi}}=
\dot\phi-{{\partial_+a}\over{\partial_-a}}\partial_-\phi=
\phi'-{2a'}{{\partial_-\phi}\over{\partial_-a}},
\end{equation}
\begin{equation}\label{pa}
P_a={{\delta L}\over{\delta\dot a}}=
{a'}\left({{\partial_-\phi}\over{\partial_-a}}\right)^2,
\end{equation}
where `dot' and `prime' denote time and spacial derivative, 
respectively.

From Eqs. (\ref{phi}), (\ref{pa}) we get the primary constraint
$$
C \equiv {1\over 4}(P_\phi-\phi')^2-P_aa'
$$
\begin{equation}\label{prima}
\equiv 
{1\over 4}(P_\phi-\phi')^2-{1\over 4}(P_a+a')^2+
{1\over 4}(P_a-a')^2 =0.
\end{equation}

The canonical Hamiltonian of the model has the form
\begin{equation}\label{ha}
H_0={1\over 2}\int dx^1(P_\phi+\phi')^2.
\end{equation}
It does not contain the field $a(x)$ and describes the energy of a 
single chiral boson mode.

The constraint (\ref{prima}) strongly commutes with $H_0$ under the 
equal--time Poisson brackets
\begin{equation}\label{pb}
\{P_\phi(x),\phi(y)\}=\delta(x-y),\qquad
\{P_a(x),a(y)\}=\delta(x-y).
\end{equation}
Hence there are no secondary constraints in the model, 
and one can check that (\ref{prima}) is the first--class constraint 
associated with the symmetry transformations (\ref{2}).
The Poisson brackets of $C(x)$ have 
the properties of a classical Virasoro stress tensor:
\begin{equation}\label{vir}
\{C(x),C(y)\}=-\delta(x-y)C'(y)+2\partial_x\delta(x-y)C(y).
\end{equation}

In contrast to the Siegel model \cite{si} where the constraint, is
\begin{equation}\label{si}
(P_\phi-\phi')^2=0, 
\end{equation}
the first--class constraint
(\ref{prima}) is not the square of a second--class constraint. 

If we partially fix the gauge under the transformations
(\ref{var}) (i.e. under $\delta a=\varphi$) by imposing the  
gauge condition
\begin{equation}\label{c2}
C_2=P_a-a'=0.
\end{equation}
we again find a relation of the present model with the McClain--Wu--Yu 
approach.

Indeed, the constraint $C_2$ is of the second class 
($\{C_2,C_2\}=2\delta'(x-y)$). And with taking it into account in 
(\ref{prima}) we can split the latter into the product of two 
multipliers
\begin{equation}\label{split}
C={1\over 4}(P_\phi-\phi'-{P_a-a'})
(P_\phi-\phi'+{P_a+a'})=0
\end{equation}
either of which can be taken as independent constraint 
(since constraints 
are always defined up to a field--dependent factor). 
For instance, let us take
\begin{equation}\label{sq}
C_1={1\over 2}(P_\phi-\phi'+{P_a+a'})=0.
\end{equation}
This constraint is still of the first class and strongly commutes 
with itself and (\ref{c2}) under the classical Poisson brackets (\ref{pb}). 

If now one would like to convert the second--class constraint into a 
first--class one by use of the standard conversion procedure \cite{fs}, 
which implies introducing new auxiliary fields, one arrives at the model 
with an infinite set of first--class constraints for an infinite set 
of fields considered in detail in \cite{mwy}. 

Let us discuss prospectives for a consistent quantization of the model 
based on action (\ref{d2}). One of the problems one should address is 
the problem of gauge symmetry anomalies. 
The indication that an anomaly might exist is the appearance of a nonzero 
central charge in the quantum commutator of constraints which are 
classically of the first class. 

In our case the quantum commutator 
acquires the central charge $c=3$ because of the sum of three 
Virasoro--like terms in (\ref{prima}).

Remember that in the Siegel model \cite{si} 
the central charge is equal to one, and to cancel the anomaly the 
authors of \cite{is} proposed to improve (\ref{si}) by adding to it 
the total derivative term $\partial^2_1\phi(x)$ with 
an appropriate coefficient. Though this way one can cancel the quantum 
anomaly, the model looses the gauge symmetry at the classical level 
since classically the new constraint is not of the first class anymore.

In our case things differ
because of the presence in (\ref{prima}) of a $b-c$ 
ghost--like term containing the auxiliary field $a(x)$.
Without spoiling the property of the constraint (\ref{prima}) to be of 
the first class we can add to it the total derivative term 
$\lambda\partial_1(P_aa)$ (where $\lambda$ is an arbitrary constant) and 
to get an improved constraint in the form
$$
C_{(\lambda)}={1\over 4}(P_\phi-\phi')^2-P_aa' 
+\lambda\partial_1(P_aa)
$$
\begin{equation}\label{improve}
={1\over 4}(P_\phi-\phi')^2-(1-\lambda)P_aa'
+\lambda P'_aa=0.
\end{equation}
This procedure is akin to ghost ``twisting'' commonly used in 
conformal field and string theory. The contribution of the terms 
containing $a(x)$ to the quantum central charge is 
$2(6\lambda^2-6\lambda+1)$ \cite{fms}. So the central charge appearing 
in the r.h.s. of the quantum commutator of (\ref{improve}) is
\begin{equation}\label{centr}
c=1+2(6\lambda^2-6\lambda+1).
\end{equation}
It vanishes at $\lambda={1\over 2}$. 

Thus we can assume that, due to operator ordering, the quantum 
theory can be reconstructed in such a way that the central charge of the 
quantum constraint (containing a contribution from ghosts (if any)) 
is equal to zero, and the anomaly associated with the local symmetry of 
the model does not arise. We hope to carry out 
detailed study of this point in future work.

\subsection{Chiral 2-forms in $D=6$}

Let us analyse from the Hamiltonian point of view the model based on 
action (\ref{fine}). (To simplify a bit the form of expressions we shall 
denote $\partial_ma(x)\equiv u_m$).
The canonical momenta of $A_{mn}$ and $a(x)$ are
$$
P_{A_{0\alpha}}=0,
$$
$$
P_{A\alpha\beta}={1\over 6}\varepsilon_{\alpha\beta\gamma\delta\rho}
F_{\gamma\delta\rho}+2{{(u_\gamma)^2}\over{u^2}}{\cal 
F}_{0\alpha\beta}
$$
\begin{equation}\label{A}
-{{u_0}\over{u^2}}
u_\gamma\varepsilon_{\alpha\beta\gamma\delta\rho}
{\cal F}_{0\delta\rho}-{2\over{u^2}} u_{[\beta}{\cal 
F}_{0\alpha]\gamma}u_\gamma
\end{equation}
$$
P_{a(x)}=2{{u_0}\over{(u^2)^2}}({u_\gamma})^2({\cal 
F}_{0\alpha\beta})^2
$$
\begin{equation}\label{ap}
-{{(u_0)^2+({u_\gamma})^2}\over{2(u^2)^2}}
u_\gamma\varepsilon_{\alpha\beta\gamma\delta\rho}
{\cal F}_{0\beta\gamma}{\cal F}_{0\delta\rho}
-4{{u_0}\over{(u^2)^2}}({\cal 
F}_{0\alpha\beta}u_\beta)^2.
\end{equation}

Remember that $\alpha, \beta,...=1,...,5$, 
$\varepsilon_{\alpha\beta\gamma\delta\rho}\equiv 
\varepsilon_{0\alpha\beta\gamma\delta\rho}$, $u^2=u_\gamma 
u_\gamma-u_0u_0$, and 
${\cal F}_{\alpha\beta\gamma}=
{1\over 2}\varepsilon_{\alpha\beta\gamma\delta\rho}
{\cal F}_{0\delta\rho}$ are the components of the anti--self--dual tensor
${\cal F}_{lmn}$ defined in (\ref{sd}).

The canonical Poisson brackets are:
$$
\{P_{A0\alpha},A_{0\beta}\}=\delta_{\alpha\beta}\delta^{(5)}(x-y),
$$
\begin{equation}\label{cpb}
\{P_{A\alpha\beta},A_{\gamma\delta}\}=
\delta_{\alpha[\gamma}\delta_{\delta]\beta}\delta^{(5)}(x-y),
\end{equation}
$$
\{P_{a}(x),a(y)\}=\delta^{(5)}(x-y).
$$
The parts of the momenta corresponding to the self--dual and 
anti--self--dual part of 
$A_{\alpha\beta}$ are
$$
\Pi^+_{\alpha\beta}=
P_{A\alpha\beta}+{1\over 6}\varepsilon_{\alpha\beta\gamma\delta\rho}
F_{\gamma\delta\rho},
$$
\begin{equation}\label{self}
\Pi_{\alpha\beta}=
P_{A\alpha\beta}-{1\over 6}\varepsilon_{\alpha\beta\gamma\delta\rho}
F_{\gamma\delta\rho},
\end{equation}
Note that $\{\Pi^+,\Pi\}=0$.

After some algebraic manipulations one gets the primary constraints:
\begin{equation}\label{ordic}
P_{A0\alpha}=0,
\end{equation}
\begin{equation}\label{varmc}
\Pi_{\alpha\beta}\partial_\beta a(x)=0,
\end{equation}
\begin{equation}\label{varc}
{1\over 8}\varepsilon_{\alpha\beta\gamma\delta\rho}\Pi_{\alpha\beta}
\Pi_{\gamma\delta}\partial_\rho a(x)+
P_a\left(\partial_\gamma a(x)\right)^2=0.
\end{equation}

The constraints are of the first class and correspond to the local 
symmetries (\ref{ordi})--(\ref{var}), respectively.

The canonical Hamiltonian is
\begin{equation}\label{hc}
H_0=\int dx^5[{1\over 
4}\left(\Pi^+_{\alpha\beta}\right)^2-
2P_{A\alpha\beta}\partial_\alpha A_{0\beta}].
\end{equation}
Commuting $H_0$ with (\ref{ordic}) we get the secondary constraint, 
which is also of the first class and corresponds to the Gauss law for 
the gauge field $A_{mn}$
\begin{equation}\label{gauss}
\partial_\alpha P_{A\alpha\beta}=0.
\end{equation}

All other constraints strongly commute with the Hamiltonian.
Thus, as in the $d=2$ case, there are no second--class constraints in the 
model, and the constraint (\ref{varc}) is not the square of a 
second--class constraint.

If we fix a Lorentz noncovariant (time--like) 
gauge $u_\gamma=\partial_\gamma a=0$, 
the definition (\ref{A}), (\ref{ap}) of the momenta implies that 
the constraints (\ref{varmc}), (\ref{varc}) consistently reduce to
\begin{equation}\label{red}
\Pi_{\alpha\beta}=0, \qquad P_a=0,
\end{equation}
where the first constraint belongs to the second class. More precisely 
it is a mixture of first-- and second--class constraints \cite{hen,bk}.
In this gauge we recover the noncovariant chiral form model of 
Refs. \cite{hen,ss}. 

\section{Conclusion}
We have demonstrated how the Lorentz covariant formulation of the 
chiral p-form model 
containing infinitely many auxiliary fields is related to the Lorentz 
covariant formulation with only one auxiliary field entering the 
chiral p--form action in a nonpolynomial way. The latter can be regarded 
as a consistent Lorentz--covariant truncation of the former.

The Hamiltonian analysis of the model based on the nonpolynomial action
has shown that in spite of nonpolynomiality the Dirac constraints have 
a simple form and are all of the first class. In contrast to the Siegel 
model the constraints are not the square of second--class constraints.
The canonical Hamiltonian is quadratic and describes a single chiral 
p--form. 

We have seen that in the case of $d=2$ chiral scalars the
constraint can be improved by use of ``twisting'' procedure 
(without the loss of the property to be of the first class) in such a 
way that the central charge of the quantum constraint algebra is zero.
This points to the possible absence  
of anomaly associated with the local 
symmetry of the classical theory in an appropriate quantum version.
To justify this conjecture one should carry out the quantization of the 
chiral form model in the formulation considered above, which is a goal 
still to be reached.

The chiral p--form action (\ref{fina}) allows coupling to gravity in the 
natural covariant way \cite{pst1}-\cite{pst3}. 
Thus the long--standing problem of 
gravitational anomaly caused by chiral forms might also be studied
in this formulation.

The nonpolynomial version can be supersymmetrized 
\cite{pst1,pst2,pst3}. In $d=2$ case an $N=1/2$  superfield formulation of
one scalar and one spinor chiral field exists \cite{pst3}, while in $D=4$ only 
a component $N=1$ supersymmetric version of duality symmetric Maxwell theory 
is known yet \cite{ss,pst2}. Recently Berkovits \cite{ber2} proposed superfield 
formulation for duality--symmetric super--Maxwell theory in the version 
with infinitely many fields. In view of the relationship considered 
above it would be of interest to truncate his supersymmetric model to a 
superfield version of the action (\ref{fine}) or (\ref{fina}).

Another interesting problem is to consider interaction of chiral forms 
with other fields and themselves
\cite{z,bs,ber2,js} with the aim, for instance, to construct complete 
actions for p-branes which have chiral form fields in their world 
volumes, such as the M--theory five--brane \cite{w,hs,js}. Our manifestly 
Lorentz covariant approach might be useful in making progress in this 
direction.

We hope to address ourselves to some of these problems in future.

\bigskip
\noindent
{\bf Acknowledgements}. The authors are grateful to I. Bandos, 
N. Berkovits,
E. Ivanov, S. Krivonos and F. Toppan for illuminating discussion.
This work was supported by the European Commission TMR 
programme ERBFMRX--CT96--045 to which P.P. and M.T. are associated.
D.S. acknowledges partial support from the INTAS Grants N 93--127 and 
N 93--493.

\end{document}